\documentclass[12pt,tightenlines,onecolumn,preprintnumbers,amsmath,amssymb,nofootinbib, prd,showpacs]{revtex4}
\usepackage{color}
\usepackage{graphicx}%
\usepackage{amsmath}
\usepackage[normalem,normalbf]{ulem}
\usepackage{amssymb}
\usepackage{bm}
\usepackage{enumerate}

\newcommand{\nn}{\nonumber}

\newcommand{\M}{\mathfrak{M}}

\newcommand{\q}{B}
\newcommand{\be}{\begin{eqnarray}}
\newcommand{\ee}{\end{eqnarray}}

\begin{document}
\title{Classification of Hypercylindrical Spacetimes with Momentum Flow}

\author{Hyeong-Chan Kim}
\email{hckim@cjnu.ac.kr}
\affiliation{School of Liberal Arts and Sciences, Chungju National University, Chungju 380-702, Korea}
\author{Gungwon Kang}
\email{gwkang@kisti.re.kr} \affiliation{Supercomputing Center, Korea
Institute of Science and Technology Information, 335 Gwahak-ro,
Yuseong-gu, Daejeon 305-806, Korea}
\author{Jungjai Lee\footnote{Corresponding authors: Jungjai Lee and Gungwon Kang}}
\email{jjlee@daejin.ac.kr}
\author{Youngone Lee}
\email{youngone@daejin.ac.kr}
\affiliation{ Department of Physics, Daejin University,
Pocheon, 487-711, Korea
}%
\date{\today}%
\bigskip
\begin{abstract}
For the five-dimensional spacetimes whose four-dimensional sections
are static, spherically symmetric ($SO(3)$) and flat asymptotically,
we study the behavior of Arnowitt-Deser-Misner mass, tension and momentum densities
characterizing such asymptotically hypercylindrical metrics under
boosts along the cylindrical axis. For such stringlike metrics two
boost-invariant quantities are found, which are a sort of ``string
rest mass-squared" and the sum of mass and tension densities.
Analogous to the case of a moving point particle, we show that the
asymptotically hypercylindrical geometries can be classified into
three types depending on the value of the ``string rest
mass-squared", namely, ``ordinary string", ``null string" and
``tachyonlike string" geometries. This asymptotic analysis shows
that the extraordinary metrics reported recently by some of the
authors belong to the tachyonlike string. Consequently, it is
likely that such extraordinary solutions are the final states of
tachyonic matter collapse. We also report two new vacuum solutions
which belong to the null string and the tachyonlike string,
respectively.
\end{abstract}
\pacs{04.70.-s, 04.50.+h, 11.27.+d, 11.25.-w}
\keywords{black hole, black string, tachyon}
\maketitle

\section{Introduction}

In general relativity with five spacetime dimensions
hypercylindrical static vacuum solutions have been studied by many
authors~\cite{Kramer,Lee,CKKL}. These solutions were extended to spacetime dimensions higher than five~\cite{kang,camps} or to inclusion of dilatonic scalar and antisymmetric form fields~\cite{Stringtheory}. (See also references therein.)

The stationary extension including a motion along the string direction was also considered by Chodos and Detweiler~\cite{Chodos} a long time ago in the context of Kaluza-Klein
dimensional reduction, and their geometrical properties were studied by Kim and Lee~\cite{kl} recently.
These stationary vacuum solutions are
the most general solutions for the metric ansatz given by
 \be
ds^2 = g_{tt} dt^2 +2g_{tz} dt dz+ g_{zz} dz^2+g_{\rho
\rho}[d\rho^2+\rho^2(d\theta^2+\sin^2\theta d\phi^2)].
 \label{Metric}
 \ee
Here the metric components are functions of the $\rho$-coordinate only.
Thus, the geometry is stationary, spherically symmetric on
any slices at $z={\rm constant}$ and uniform along the $z$-direction.

Such solutions are characterized by three parameters. By
considering the asymptotic behaviors of the metric components at
infinity, one can see that these parameters correspond to ADM mass, tension and momentum densities. The presence of the additional momentum density parameter appears to be due to a ``constant'' motion of the stringlike object~\cite{Chodos,kl}.
Therefore, since the geometry is
stationary and uniform along the $z$-direction, one might expect
that the non-vanishing momentum density could be removed by a
suitable boost transformation along that direction. Namely, one may
consider an observer moving along the string direction at the same
speed. Interestingly, however, it was noticed that the momentum
parameter is not always removed by boost
transformations~\cite{Chodos,kl}.
Thus, it is not understood well how such spacetime could be formed.
Moreover, it has not been studied well about what values of those three parameters are allowed physically.

In this paper, we investigate the behavior of asymptotic ADM
quantities characterizing hypercylindrical spacetimes under arbitrary boost
transformations along the $z$-direction in order to understand such
interesting behavior better.
In the case of a point particle moving along the $z$-direction with energy
$E$ and momentum $p_z$ in the Minkowski spacetime, although both energy and
momentum are not invariant under a boost, one can find a boost invariant combination of them. This is nothing but the rest mass-squared, {\it i.e.}, $m_0^2\equiv  E^2 -p_z^2$.
Depending on the sign of the rest mass-squared, the particles are divided into three classes, {\it i.e.,}``ordinary particle" ($m_0^2>0$), ``null particle" ($m_0^2=0$) and ``tachyon" ($m_0^2<0$), respectively.
In the case of a uniform stringlike object moving along the string direction, on the other hand, neither such invariant quantity nor classification is known.

Based on the behavior of asymptotic ADM quantities
under boost transformations we define a string rest mass-squared which plays the same role as the particle rest mass-squared.
We also show that this definition is related to the dominant energy condition applied to hypercylindrical matter distributions.
In analogy with a moving particle case, the asymptotic geometries may also
be divided into three types, namely, "ordinary string", "null string" and "tachyonlike string".
This analysis on the ADM parameters on boost transformations indicates that the extraordinary solution in Ref.~\cite{kl} belongs to the tachyonlike string geometry.
Consequently, it is likely that such an extraordinary solution is a final state of tachyonic matter collapse.
In fact, tachyon scalar field of an unstable D-brane or a brane-antibrane system are of very interest in the context of string theory and tachyon cosmological models~\cite{string cosmology}. (See also references therein.)

In Sec.~\ref{app:Econd}, by considering the dominant energy condition to see if it gives some constraints on the physical values of the ADM parameters, the "gravitational dominant energy condition'' is imposed.
In Sec.~\ref{class}, we study the behaviors of asymptotic forms of a stringlike
geometry under boost transformations along the string.
We also classify these asymptotic spacetimes according to their transformation properties.
In Sec.~\ref{classofsolutions}, we explicitly present several exact vacuum solutions belonging to those classes. In Sec.~\ref{summary}, we summarize the results and discuss physical implications of them.

\section{Physical ranges in ADM parameters}
\label{app:Econd}

Consider any asymptotically hypercylindrical spacetimes which are
described by the metric in the form of Eq.~(\ref{Metric}). By
asymptotically hypercylindrical metrics we mean, at least at spatial
infinity, that the four-dimensional section of the five-dimensional
spacetime under consideration is static, spherically symmetric
($SO(3)$) and flat, and that the geometry is uniform along the fifth
direction with also admitting a constant momentum flow. Then the
asymptotic forms of such spacetime metrics may be given by
 \begin{eqnarray}
  \label{metric:comp}
g_{tt}&\simeq& -1+\frac{4G(2M-\tau)}{3\rho},~~
g_{zz} \simeq 1+\frac{4G(M-2\tau)}{3\rho}, \\
 g_{tz}&\simeq& -\frac{4G P}{\rho},
 ~~~ ~~~~~~~~~~~~~~~ g_{\rho\rho} \simeq 1+\frac{4G(M+\tau) }{3 \rho}\, ,\nn
 \end{eqnarray}
where $G$ is  the 5-dimensional Newton's constant. These spacetimes
could arise from a matter source localized in 3-spatial dimensions
which is uniformly extended to and moving along the $z$-direction.
By applying the definitions of gravitational ADM quantities, one can
see that $M$, $\tau$ and $P$ above indeed denote the ADM mass,
tension and momentum densities, respectively.

Even if ranges of these three integration constants are arbitrary mathematically, physical values may be restricted.
It is well known that the ADM mass of an asymptotically flat spacetime is non-negative provided that the dominant energy condition is satisfied.
It is also proved that the purely gravitational contribution to the spacetime tension is positive definite for transverse asymptotically flat spacetimes without horizons that arise from brane matter sources~\cite{traschen04}.
We expect that the magnitude of the gravitational momentum $P$ may be limited by given values of $M$ and $\tau$.
Finding such physical ranges in gravitational ADM quantities would be very difficult in general.

In order to get some hints, let us consider ordinary matter distributed  hyper-cylindrically in five-dimensional flat spacetime.
Let the matter distribution take the form of a moving fluid with tension $\tilde \tau$ and momentum $\tilde p$ along $z$:
\begin{eqnarray} \label{T}
T_{ab} =\tilde\rho u_a u_b -\tilde\tau z_az_b+ \tilde p(u_a z_{b}+ z_{a} u_b) ,
\end{eqnarray}
where $u_a$ and $z_{a}$ denote the timelike and spacelike Killing
vectors orthonormal to each other so that the matter distribution
possesses the translational symmetry along $z$. Here we assumed, for
simplicity, that the matter distribution has a spatial translational
symmetry along the $x^4=z$ direction with $\tilde\tau_{i}=0=\tilde
p_i$ for $i=1,2,3$ and has vanishing shears.

Now we consider the dominant energy condition for matter, that is, $j_a=-T_{ab}\xi^b$ is a future directed timelike vector for any timelike vector $\xi^a$.
Without loss of generality, we may choose $\xi^a = u^a+ \beta z^a $ with $|\beta|<1$.
The dominant energy condition restricts the energy, tension, and momentum densities.
The timelikeness of $j_a$ restricts
\begin{eqnarray} \label{energycond}
\tilde\rho - \tilde\tau \geq 2|\tilde p|, \quad \quad
\end{eqnarray}
and its future directedness gives
\begin{eqnarray}
\tilde\rho+\tilde\tau \geq 0 .
\end{eqnarray}
Thus we can see that the energy density $\tilde\rho$ is nonnegative and that the magnitudes of the momentum and tension densities cannot be bigger than the energy density, {\it i.e.,} $|\tilde p|,~ |\tilde\tau|\leq \tilde\rho$.

Suppose that this hypercylindrical matter distribution is confined
at the central region. Then, in the linearized gravity analogy, it
is known that the ADM mass, tension and momentum densities at
transverse asymptotically flat region are given by $ M=\int
dV\,\tilde\rho $, $ \tau =\int dV\, \tilde\tau$ and $P = \int dV \,
\tilde p$, respectively. Here $dV$ is the spatial volume element at
the $z=$ constant surface. Thus, at least for hypercylindrical
matter distributions producing weak gravitational fields around, we
see that the ADM gravitational quantities satisfy
 \begin{eqnarray} \label{energycond2}
\frac{M-\tau}2 \geq |P|, \quad \quad M+\tau \geq 0.
 \end{eqnarray}
Consequently, this condition above gives several bounds for the ADM
quantities. Namely, the ADM mass is non-negative ($M \ge 0$), which
is consistent with the positive energy theorem if it holds even in
our case, and the physical values of $P$ and $\tau$ are not
arbitrarily large but bounded such that
$$ -M \leq P,~ \tau \leq M $$
for a given value of the ADM mass density. We do not know that these
conditions will still be satisfied in general. However, it is likely
that the ADM quantities also satisfy Eq.~(\ref{energycond2}) if
hypercylindrical spacetime solutions are the end state of ordinary
matter collapse satisfying the dominant energy condition. Therefore,
we conjecture a sort of gravitational dominant energy condition
that the ADM mass, tension and momentum densities of a transverse
asymptotically flat stringlike spacetime satisfy
Eq.~(\ref{energycond2}) provided that the dominant energy condition
for matter is satisfied.

\section{Classification of Asymptotic String-like Geometries}
\label{class}

Let us consider an observer $\mathcal{O}'$ moving along the
$z$-direction with a constant velocity $v_z=\tanh \xi$ with respect
to an observer $\mathcal{O}$ in $(t,z)$~(\ref{metric:comp}). The coordinates $(t',z')$
for $\mathcal{O}'$ can be obtained from the boost transformation
given by
\begin{eqnarray} \label{eq:boost:asym}
\left(\begin{tabular}{c}
    ${t'}$\\ ${z'}$ \\\end{tabular}\right)
    &=&\left(\begin{tabular}{c}
    $\cosh \xi~~ -\sinh\xi$\\ $-\sinh\xi~~
        \cosh\xi$ \\ \end{tabular}\right)
\left(\begin{tabular}{c}
    $t$\\ $z$ \\\end{tabular}\right).
\end{eqnarray}
Consequently, the ADM mass, tension and momentum densities in the primed
coordinates are given by
 \begin{eqnarray}
\label{p:np}
M' &=& M \cosh^2\xi -\tau \sinh^2\xi -P \sinh 2\xi,  \nn\\
\tau' &=& -M \sinh^2\xi +\tau \cosh^2\xi +P \sinh 2\xi, \nn\\
P' &=& -\frac{1}{2}\left( M-\tau \right) \sinh 2\xi +P \cosh 2\xi.
 \end{eqnarray}

As in the case of a moving particle, we expect that there are some boost invariant quantities.
One can easily see that the sum of the mass and
tension densities is one of such quantities.
For later convenience we define a boost invariant quantity
 \begin{equation}
 \label{M}
\M \equiv M+\tau.
 \end{equation}
The invariance of $\M$ can be understood in the following way: For
the case of matter in Eq.~(\ref{T}),  the trace of the stress-energy
tensor is $T^a_a = -(\tilde\rho+\tilde\tau)$.~\footnote{We point out
that the actual trace has additional contributions coming from the
pressure on the perpendicular plane to the $z$-direction which we
ignored for simplicity. Since those terms are scalars under
$z$-boosts, the considerations below are not affected. However, it
should be pointed out that the tracelessness does not imply
$\tilde{\tau} = -\tilde{\rho}$ for actual matters.} Note that this
quantity is scalar which is invariant under arbitrary coordinate
transformations at any spacetime point. Then, we can easily see that
the integrated quantity
$$-\int dV\, T^a_a = \int dV\,( \tilde\rho +\tilde \tau) $$
is also invariant under the $z$-boost.
Here the spatial volume $dV$ is orthogonal to the $z$ direction.
In the linearized gravity analogy,
therefore, the boost invariance of this integral for matter implies the boost invariance of $\M=M+\tau$ subsequently.

Note that
$$
\frac{1}{2} T_{ab} T^{ab}-\frac{1}{4} (T^a_a)^2 = \frac{(\tilde\rho
-\tilde\tau)^2}{4}-  \tilde p^2
$$
is also a scalar quantity. Although any combination of $T_{ab}
T^{ab}$ and $(T^a_a)^2$ becomes a scalar quantity, this combination
above is special in the following sense. Namely, only this quantity
changes its sign from positive to negative when the dominant energy
condition is violated. Since the dominant energy condition implies
that the trajectory of physical matter is timelike or null, the
violation of such condition indicates the crossing of the light
cone. In the linearized gravity analogy, the corresponding scalar
quantity is given by
 \begin{eqnarray}
 \label{M0}
M_0^2 \equiv  \frac{(M-\tau)^2}{4} - P^2.
 \end{eqnarray}
Indeed it can be shown from Eq.~(\ref{p:np}) that this quantity is boost invariant.

The quantity $M_0^2$ is a
sort of  ``string rest mass-squared'' analogous to the rest mass-squared of a moving
particle.
Note that $M^2 -P^2$, which is analogous to the rest mass-squared for a moving particle, is not boost invariant in the case of
stringlike objects. Note also that $M_0^2$ is not positive definite
in general.

It turns out that the following combination is useful:
\begin{equation} \label{q}
\q_\pm  \equiv \frac{M-\tau}{2}\mp P.
\end{equation}
Note that $\q_+\q_-=M_0^2$.
These quantities are observer dependent, but transform in a simple way as
\begin{equation} \label{boost:q}
\q_\pm' = e^{\pm 2\xi}\q_\pm
\end{equation}
under the boost. Note that the sign of $\q_\pm $ does not change.
For the case of a particle $B_\pm$ corresponds to $E\mp p_z$.
The definiteness of the sign of $E\mp p_z$ guarantees causality. Namely, a timelike (spacelike) motion remains timelike (spacelike) under boost transformations.

In the new variables, the gravitational dominant energy condition~(\ref{energycond2}) can be expressed as
\begin{equation}
M_0^2\geq 0, \qquad \M \geq 0, \qquad B_+ \geq 0. \label{DEC}
\end{equation}
If $\q_\pm \neq 0$, the ADM mass, tension, and momentum densities can be written in terms of the observer invariant and dependent quantities by:
\begin{eqnarray} \label{MtP}
M  &=&  \frac12\left(\M +\q_\pm +\frac{M_0^2}{\q_\pm } \right), \nn \\
\tau &=&   \frac12\left(\M-\q_\pm -\frac{M_0^2}{\q_\pm }  \right), \nn \\
\quad P &=& \pm \frac{1}{2}\left(-\q_\pm +\frac{M_0^2}{\q_\pm }  \right).
\end{eqnarray}
The set of ADM quantities $(\M, M_0^2,\q_\pm)$ has one to one relation with the other set of ADM quantities $(M,\tau,P)$ provided that $B_\pm \neq 0$.
The case of $B_\pm =0$ will be treated below.

\begin{figure}[htb]
  \begin{center}
 \includegraphics[width=.5\linewidth]{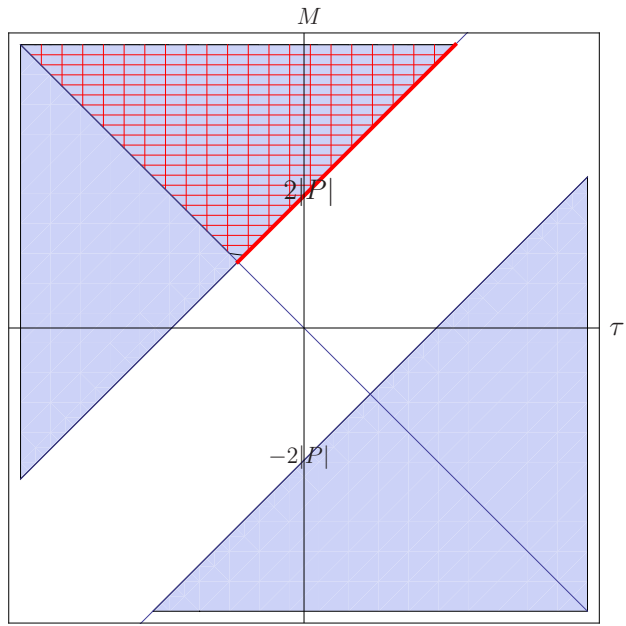}
  \end{center}
  \caption{Classification of string solutions.
  The solid lines are given by $M= \tau \pm 2|P|$ and $M=-\tau$.
  The red-meshed region satisfy the dominant energy conditions.
  }\label{fig:Econd}
\end{figure}

It is convenient to divide the ADM parameter space into three cases depending on the sign of $M_0^2$ as in Fig.~\ref{fig:Econd}.
We call these cases as ordinary string, null string and tachyonlike string for $M_0^2>0$, $M_0^2=0$ and $M_0^2<0$, respectively.
Note that $M_0^2=0$ is the boundary for the energy condition in Eq.~(\ref{DEC}) to be satisfied.
The behaviors of the mass and momentum densities under the boost are explicitly shown in
Fig.~\ref{fig:ADM}.\footnote{Although this figure shows the relationship between $M$ and $\q_+$ or between $P$ and $\q_+$, we can also interpret this as the behaviors of $M$ and $P$ under boost by regarding $\q_+$ as a boost parameter since $\q_+' = e^{2\xi} \,\q_+$ and $\M'=\M$.    } The variation of tension is readily readable since $\M=M+\tau$ is boost invariant.

\begin{itemize}
\item Ordinary string  [$M_0^2>0$]; This class corresponds to the shaded regions in Fig.~\ref{fig:Econd}, and is the counterpart of a moving particle having timelike trajectory.
The ``string energy condition"~(\ref{energycond2}) is satisfied for the case of the meshed region only. The (boosted) Schwarzschild black string belongs to this case.
In this meshed region, as shown in the black dashed and dotted curves in Fig~\ref{fig:ADM}, the momentum value monotonically decreases if an observer moves parallel to the initial string momentum direction.
As the velocity of the observer increases, it crosses zero at
$|\q_+|=\sqrt{M_0^2}$ for an observer who comoves with the string object.
The ADM mass on the other hand decreases as the boost increases and bounces up after passing a minimum value for the comoving observer.
The ADM tension, on the other hand, increases as the boost increases and bounces down after passing a maximum value.
In other words, the ADM tension takes maximum value when the relative velocity between the observer and the string vanishes.
As the relative velocity increases, the ADM tension decreases gradually.

\begin{figure}[htb]
  \begin{center}
\begin{tabular}{cc}
 \includegraphics[width=.53\linewidth]{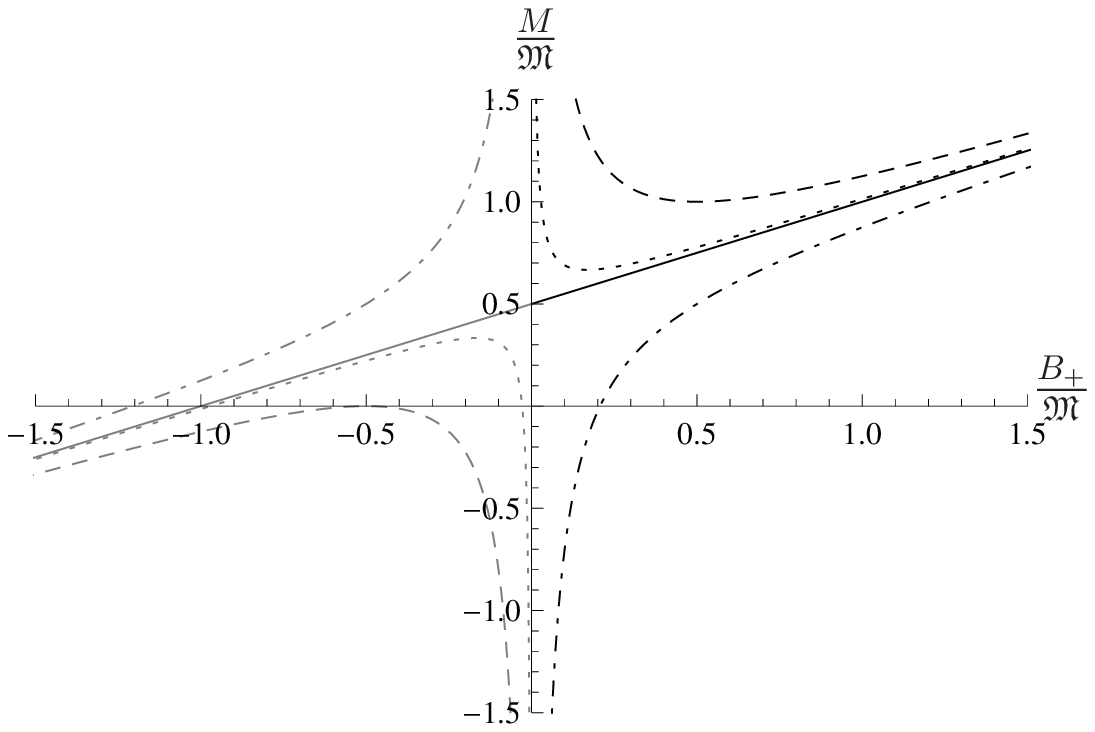} &
      \includegraphics[width=.53\linewidth]{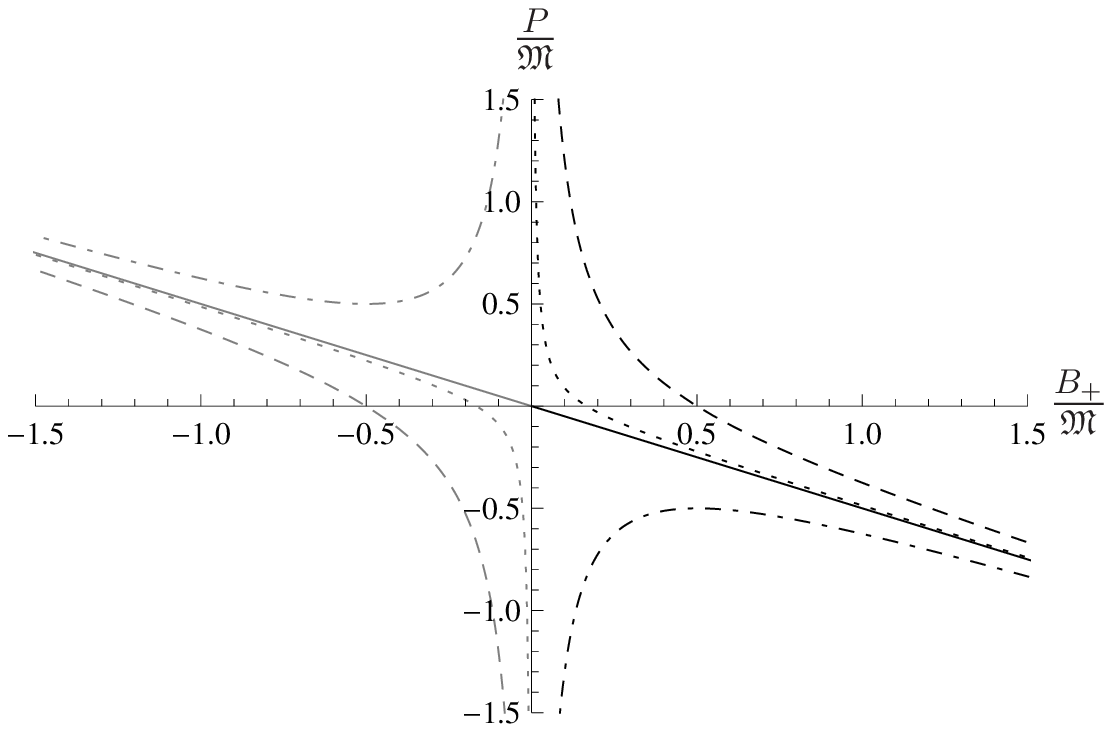}
 \end{tabular}
  \end{center}
  \caption{
  The change of the ADM mass and momentum densities for ($\q_+/\M$) for several different values of string rest mass-squared.
  The dashed, dotted, black (gray), and dot-dashed curves correspond to $M_0^2/\M^2=  1/4, 1/36,0$, and $-1/4$, respectively.
The boosted Schwarzschild black string and the Kaluza-Klein bubble solution correspond to the black dotted curve and the gray dotted curve with $M_0^2/\M^2= 1/36$, respectively.
   }\label{fig:ADM}
\end{figure}

\item Null string [$M_0^2=0$]; This class consists of the boundary lines between the shaded regions and the unshaded region in Fig.~\ref{fig:Econd}, and is the counterpart of a massless particle.
The condition~(\ref{energycond2}) is satisfied for the thick line only.
The change of the mass and momentum densities under the boost are plotted as
the black and gray solid lines in Fig.~\ref{fig:ADM} which correspond to the case $B_-=0$.

Since $\q_+\q_- = M_0^2$, the case of $M_0^2=0$ ({\it i.e.,} $M-\tau=\pm 2P$) gives $B_+ =0$ or $B_-=0$.
In case of $B_+=0$,
we may use the set $(\M, M_0^2, \q_-)$ in Eq.~(\ref{MtP}).\footnote{
If $B_-=0$, {\it vice versa}.
In case that both $B_+=0=B_-$, we have $P=0$ and $M=\tau=\M/2$.}
Namely, the mass, tension and momentum densities are given by
$$
M = \frac{\M +\q_-}2, \quad \tau=\frac{\M-\q_-}2, \quad P = \frac{\q_-}2.
$$

The boost transformation~(\ref{boost:q}) leads to
\begin{equation} \label{case2}
P'= P e^{-2\xi}.
\end{equation}
Note that the sign of the momentum does not change under the motion of observer and we cannot make the momentum vanish by boosts if the initial momentum is nonzero.
In addition, if the initial momentum is zero, we cannot make it be finite by using the $z$-boost.
This property is similar to the case of chasing a massless particle.

\item Tachyonlike string  [$M_0^2< 0$];  This class consists of the white region in Fig.~\ref{fig:Econd}.
The black dot-dashed and gray dot-dashed curves in Fig.~\ref{fig:ADM} denote the change of the mass and momentum densities under the boost.
For given values of $M_0^2$ and $\M$,  Fig.~\ref{fig:ADM} shows that the ADM momentum squared has a nonvanishing minimum value given by
$$
P^2=-M_0^2.
$$
This point can always be reached by a suitable boost transformation, and we have $M=\tau$ and $B_+ = |M_0|$ or $-|M_0|$ in this case.
Note that any other observer measures $M\neq \tau$, resulting in the growth of the momentum squared given by the following formula:
$$
P^2 = -M_0^2 + \frac{(M-\tau)^2}{4} .
$$
Notice also that the mass monotonically changes under the boost.

\end{itemize}

So far we have studied the behavior of ADM quantities under the boost transformations and subsequent classifications of the stringlike spacetime geometries.
Here we show how the geodesic motions in the tachyonlike string differ from those in the ordinary string at the asymptotic region.
The effective potential for the radial motion of a geodesic in the spacetime with metric~(\ref{Metric}) is given by
$$
V_{\rm eff}(\rho) =
\frac{1}{g_{\rho\rho}}\left(g^{tt}E^2-2Ep_z g^{tz}+p_z^2 g^{zz} +L^2g^{\theta\theta} +m^2 \right),
$$
where $L$, $E$, and $p_z$ are the spatial angular momentum around the three dimensions, the energy, and the momentum along $z$ coordinate of the test particle, respectively.
The test mass-squared $m^2$ is positive, zero and negative depending on timelike, null and spacelike geodesics, respectively.

The asymptotic behavior of this effective potential becomes
\begin{eqnarray*}
V_{\rm eff}(\rho) &\simeq& -p_\perp^2 -
 \left[(m^2-p_\perp^2)\M
    +\frac{3M_0^2}{\q_+} (E-p_z)^2 +3 \q_+(E+p_z)^2\right] \frac{2G}{3\rho} ,
\end{eqnarray*}
where $p_\perp^2= E^2-p_z^2-m^2$ is the momentum density of the test particle at $\rho=$ infinity along the direction orthogonal to the string.

Let us consider the motion of a test particle  with $p_\perp=0$ moving on the background of string solutions satisfying gravitational dominant energy condition, {\it i.e.,} $\M>0$, $B_+>0$, and $M_0^2>0$.
Then the form of the effective potential shows that all test particles are attracted toward the string center.
On the tachyonlike string background with $M_0^2<0$ keeping $\M>0$ and $B_+>0$, on the other hand, some test particles are repelled by the string even in the absence of the angular momentum.

In this sense, the geometric property of the tachyonlike string is critically different from that of the ordinary string.

\section{Vacuum String Solutions }
\label{classofsolutions}

In this section we explicitly demonstrate some exact vacuum solutions which belong to each case in the previous section.
The most general vacuum solution to the Einstein equation with the ansatz~(\ref{Metric}) are as follows:
\begin{eqnarray} \label{Sol}
g_{tt}&=&
D^{-\frac{2}{\sqrt{3}\sqrt{1+\chi^2}}}
    \left(s \,D^{\frac{2\chi}{\sqrt{1+\chi^2}}}-c\, D^{-\frac{2\chi}{\sqrt{1+\chi^2}}}
    \right),\\
g_{zz}&=&
    D^{-\frac{2}{\sqrt{3}\sqrt{1+\chi^2}}}
    \left(c \, D^{\frac{2\chi}{\sqrt{1+\chi^2}}}-
    s \,D^{-\frac{2\chi}{\sqrt{1+\chi^2}}}\right) \nn ,\\
g_{tz} &=&-(cs)^{1/2} \,D^{-\frac{2}{\sqrt{3}\sqrt{1+\chi^2}}}
    \left(D^{\frac{2\chi}{\sqrt{1+\chi^2}}}-
     D^{-\frac{2\chi}{\sqrt{1+\chi^2}}}\right),\nn \\
g_{\rho\rho}&=&\left(1-\frac{K^2}{\rho^2}\right)^{2}
    D^{\frac{4}{\sqrt{3}
        \sqrt{1+\chi^2}}} , \nn
\end{eqnarray}
where $c-s=1$ and $ D(\rho) = (1+K/\rho)/(1-K/\rho)$.
Here the parameters  $c,s,K$ and $\chi$ are complex numbers in general and it turns out that the following values of parameters make real metric components:
\begin{eqnarray} \label{classify}
\mbox{class I}: c&=& \cosh^2 \xi, ~~ s= \sinh^2\xi, ~~ \xi,\chi, K \in R \,, \\
\mbox{class II}: c&=&\frac{1}{2}- i q,~~s=-\frac{1}{2}-i q,~~\chi = -i \bar
    \chi; ~|\bar\chi|\leq 1,\quad q,\bar \chi, K\in R \,, \label{class2}\\
\mbox{class III}: c&=&\frac{1}{2}- i q,~~s=-\frac{1}{2}-i q,~~\chi = -i
    \bar \chi; ~|\bar \chi|\geq 1,~~K= i Q, \quad q, \bar \chi,Q\in R \label{class3} .
\end{eqnarray}
These solutions were, in fact, found by Chodos and Detweiler~\cite{Chodos} and
later analyzed in Refs.~\cite{kl,Gwak}.
The ADM quantities are represented by
\begin{eqnarray}
M= \frac{\sqrt{3}+(c+s)\chi }{\sqrt{1+\chi^2}} K, \quad
\tau =  \frac{\sqrt{3}-(c+s)\chi }{\sqrt{1+\chi^2}} K  , \quad
P = \frac{2\sqrt{cs}\,\chi}{\sqrt{1+\chi^2}} K.  \label{ADM:gen}
\end{eqnarray}
Then, $M_0^2$, $\M$ and $\q_\pm$ are given by
\begin{eqnarray}
M_0^2 =\frac{\chi^2}{1+\chi^2}K^2, \quad
\M = \frac{2\sqrt{3}}{\sqrt{1+\chi^2}} K, \quad
\q_\pm=\frac{(c^{1/2}\mp s ^{1/2})^2\chi}{\sqrt{1+\chi^2}}K.
\label{Inv:gen}
\end{eqnarray}
Thus we can see that $\chi$ and $K$ are boost invariant parameters whereas $c$ (or $q$) is boost dependent.
Note also that $M_0^2\geq 0$ for class I and $M_0^2\leq 0$ for class II and III.
Since $-M_0^2/\M^2= \bar \chi^2/12$, we have $0\leq -M_0^2/\M^2\leq /12$ for class II and $-M_0^2/\M^2\geq /12$ for class III.

\subsection{Ordinary string solutions}

In this subsection, we consider the solutions whose asymptotic parameters satisfy $M_0^2> 0$. The class I solutions in Eq.~(\ref{classify}) except for the case of $\chi=0$ belongs to the ordinary string.
As explained in the previous section, we can make any finite value of ADM momentum parameter be zero by a suitable boost transformation in this case.
Thus we consider only the case $\xi=0$ ({\it i.e.,} $c=1$ and $s=0$).

The solutions in this case are static and were found in Refs.~\cite{Kramer,Chodos}.
The metric is given by~\cite{Lee,CKKL}
\begin{eqnarray} \label{lee}
ds^2 &=&-\left(\frac{1+K/\rho}{1-K/\rho}\right)^{-\frac{2(1+\sqrt3 \chi)}{\sqrt{3(1+\chi^2)}}}
    dt^2+
    \left(\frac{1+K/\rho}{1-K/\rho}\right)^{-\frac{2(1-\sqrt3 \chi)}{\sqrt{3(1+\chi^2)}}}
    dz^2  \\
&+&\left(1-\frac{K^2}{\rho^2}\right)^2
    \left(\frac{1+K/\rho}{1-K/\rho}\right)^{\frac{4}{\sqrt{3}
        \sqrt{1+\chi^2}}}
    \left(d\rho^2+
    \rho^2 d\Omega_{(2)}^2\right). \nn
\end{eqnarray}
The metric~(\ref{lee})  becomes the well-known Schwarzschild black string solution when $\chi=1/\sqrt{3}$ with positive $K$.
For this case we see that $M_0^2=K^2/4$, $\M=3K$, and $B_+=  K/2$, satisfying the gravitational dominant energy condition in Eq.~(\ref{DEC}).

The Kaluza-Klein bubble solution, which can be obtained by the double-Wick rotation $t\to i z$ and $z\to i t$, corresponds to $\chi=-1/\sqrt{3}$ with positive $K$. Therefore, we have $M_0^2=K^2/4$, $\M=3K$, and $B_+=  -K/2$.
Thus this case does not satisfy the gravitational dominant energy condition.
Consequently, the Kaluza-Klein bubble metric cannot presumably be formed as a final state of ordinary matter collapse satisfying dominant energy condition.
Then, one may speculate it is formed as a final state of the collapse of past directed matter.
However, it is not true because the future-directed condition is satisfied, $\M>0$.
The geodesic motions in this spacetime were studied in Ref.~\cite{Gwak} in detail.

\subsection{Null string solutions}
This case corresponds to $M_0^2=0$ or $M-\tau=\pm 2P$.
$M_0^2=0$ gives $K=0$ or $\chi=0$.
The case of $K=0$ gives the flat spacetime which we are not interested in.
In Eq.~(\ref{ADM:gen}), we may have $c^{1/2} \to \infty$ and $s^{1/2}\to \pm \infty$ as $\chi \to +0$ to give a finite nonvanishing momentum density.
Thus, by taking the limit of $\pm (cs)^{1/2}\to \infty$ and $\chi \to +0$ with $\alpha= \sqrt{cs}\,2\chi /\sqrt{1+\chi^2}$ fixed, we have
\begin{eqnarray}
ds^2 &=& \left(\frac{1+K/\rho}{1-K/\rho}\right)^{-\frac{2}{\sqrt{3}}}
    \left[-dt^2 + dz^2 +2\alpha \log\frac{\rho+K}{\rho-K} (dt\mp dz)^2\right] \nn \\
    &+& \left(1-\frac{K^2}{\rho^2}\right)^2 \left(\frac{1+K/\rho}{1-K/\rho}\right)^{\frac{4}{\sqrt{3}}}
    \left(d\rho^2 +\rho^2 d\Omega_{(2)}^2 \right).  \label{sol2}
\end{eqnarray}
We can directly check that the above metric is a solution of the Einstein equation.
This metric with nonvanishing $\alpha$ has not been reported in the literature as far as we know.

Note that the ADM parameters are given by
$$
M = \left(\sqrt{3}+\alpha \right)K, \quad \tau =  \left(\sqrt{3}-\alpha \right)K, \quad P = \pm \alpha K.
$$
We see that the three-dimensional area of $\rho=\mbox{constant}$ surface becomes infinite as $\rho \to K$ and that this surface is indeed a naked singularity.

\subsection{Tachyon-like string solutions}

The ADM parameters satisfy $M_0^2 < 0$.
In this case, as explained above, we can always make a suitable boost transformation such that $M=\tau$ with $\q_+= |M_0|$ or $-|M_0|$.
This choice determines $c=1/2=-s $ and the solutions are characterized by $\chi$ and $K$ in Eqs.~(\ref{class2}) and (\ref{class3}). Here $\chi$ can be expressed as $\chi= -i \sqrt{3}\mathfrak{p}$ where $\mathfrak{p}= P/M$ is the momentum to mass ratio.

Now the metric for this case is expressed as~\cite{kl}
\begin{eqnarray}
ds^2&=& f^{-1}(\rho) \left[-(\omega^t)^2+(\omega^z)^2\right] +
f^2(\rho) g^2(\rho) (d\rho^2+\rho^2
d\Omega_{(2)}). \label{ds2:extra}
\end{eqnarray}
Here the timelike 1-form $\omega^t$ and the spacelike 1-form $\omega^z$ are
\begin{eqnarray} \label{eq:D}
\omega^t = \cos(\mathfrak{p}\Upsilon) \, dt - \sin(\mathfrak{p}\Upsilon) \,dz, \quad \omega^z
= \sin(\mathfrak{p}\Upsilon) \,dt +\cos(\mathfrak{p}\Upsilon) \, dz,
\end{eqnarray}
with Killing coordinates $t$ and $z$.
For $|\mathfrak{p}|< 1/\sqrt{3}$ (class II),\footnote{The case of $\mathfrak{p}=\pm 1/\sqrt{3}$ will separately be treated below } the functions $f(\rho)$, $g(\rho)$ and $\Upsilon(\rho)$
are given by
\begin{eqnarray}
f(\rho) =\left(\frac{1+K/\rho}{1-K/\rho}\right)^{\frac{2}{\sqrt{3}\sqrt{1-3\mathfrak{p}^2}}} , \quad
g(\rho)=1-\frac{K^2}{\rho^2}, \quad \Upsilon(\rho)=\frac{\sqrt{3}}{\sqrt{1-3\mathfrak{p}^2}}
\,\log\frac{1+K/\rho}{1-K/\rho}, \label{sol:class2}
\end{eqnarray}
Note that, as $\mathfrak{p} \to 0$ ({\it i.e.,} $\chi\to 0$), this solution becomes the null string solution~(\ref{sol2}) with $\alpha=0$.
The function $\Upsilon$ monotonically increases from zero to infinity as $\rho$
decreases from infinity to $K$.
The geodesic motions in this spacetime are studied in Ref.~\cite{bo}.
At $\rho=K$, except for the case of $|\mathfrak{p}| = \sqrt{5/27}$,  there exists a curvature singularity which is naked for $|\mathfrak{p}|< 1/(2\sqrt{3})$ and null for $1/(2\sqrt{3})\leq |\mathfrak{p}| < \sqrt{2}/{3}$.
There is no curvature singularity if $|\mathfrak{p}| \geq \sqrt{2}/{3}$.

For $|\mathfrak{p}|> 1/\sqrt{3}$ (class III), we have
\begin{eqnarray}
f(\rho) = \exp\left[\frac{2\tan^{-1} (Q/\rho)}{\sqrt{3}\sqrt{3\mathfrak{p}^2-1}}\right], \quad
g(\rho)= 1+\frac{Q^2}{\rho^2}, \quad
\Upsilon(\rho) = \frac{\sqrt{3}\tan^{-1}(Q/\rho)}{\sqrt{3 \mathfrak{p}^2 -1}} ,
\end{eqnarray}
which describe wormhole solutions regular everywhere~\cite{Chodos,kl}.
The function $\Upsilon$ monotonically increases from zero to $\pi/2$ as $\rho$ decreases from infinity to zero.

For the case of $|\mathfrak{p}|= 1/\sqrt{3}$ ({\it i.e.,} $\chi^2=-1$), the ADM quantities and the metric components appear to be singular.
However, regular solutions exist when $K\to 0$ as well with keeping $ \lambda= \frac{2K}{\sqrt{3}\sqrt{1-3\mathfrak{p}^2}}$ be fixed.
\begin{eqnarray}\label{sol4}
f(\rho) = e^{2\lambda/\rho}, \quad g(\rho) =1, \quad \Upsilon(\rho) = \frac{3\lambda}{ \rho} .
\end{eqnarray}
The ADM quantities become
$$
M=\tau= \frac{3\lambda}{2}, \quad P=\pm \frac{\sqrt{3}}{2}\lambda .
$$

\section{Summary and discussion}
\label{summary}

We have considered asymptotically stationary hypercylindrical metrics in five dimensions characterized by the ADM mass, tension and momentum densities.
Motivated by the linearized gravity analogy,  the gravitational dominant energy condition was conjectured by which the physical ranges of these three quantities are restricted as in Eq.~(\ref{energycond2}).
The study of boost transformations of these three quantities for such stringlike metrics along the string shows that there are two boost-invariant quantities, which are the string rest mass-squared ($M_0^2$) and the ADM (mass$+$tension) density ($\M$).
Analogous to the case of a moving point particle, we show that the asymptotically hypercylindrical geometries can be classified into three types depending on the value of the string rest mass-squared, namely, ordinary string ($M_0^2>0$), null string ($M_0^2=0$) and tachyonlike string ($M_0^2<0$) geometries.
Note that the hypercylindrical solutions satisfying the gravitational dominant energy condition such as the Schwarzschild black strings belong to the ordinary string.
The Kaluza-Klein bubble solution also belongs to the ordinary string, but it does not satisfy the gravitational dominant energy condition.

This analysis on boost transformations for ADM parameters shows that the extraordinary solutions ({\it e.g.,} class II in Eq.~(\ref{class2})) reported in Refs.~\cite{Chodos,kl} belong to the tachyonlike string.
We expect that the collapse of hypercylindrical matter distribution with momentum flow ends up with stationary vacuum solutions in Eq.~(\ref{Sol}).
If the collapsing matter satisfies the dominant energy condition, all the way down to the final stationary state, the spacetime produced will probably belong to the ordinary string with the gravitational dominant energy condition satisfied.
Although the details of the collapsing processes are not known, we expect that the collapse of tachyonic matter presumably ends up with some stationary spacetimes belonging to the tachyonlike string, for instance, the extraordinary solutions.

In this paper, we have not given any restriction on the values of three ADM quantities asymptotically characterizing the hypercylindrical spacetimes with momentum flow.
Usually, one imposes some conditions on these ADM quantities such as the gravitational dominant energy condition we conjectured and simply discards all other solutions which do not satisfy such conditions since those solutions are probably resulted from the collapse of unphysical matter.
The reasons we consider those solutions that do not satisfy the gravitational dominant energy condition as well are the following:
First of all, notice that the well known Kaluza-Klein bubble solution does not satisfy the gravitational dominant energy condition although it belongs to the ordinary string.
However, this solution does not have a naked singularity and is perfectly regular everywhere provided that the $z$ coordinate is suitably compactified.\footnote{
Note that a boost along the $z$-direction at $\rho=\infty$ is still a constant motion even if the $z$ is an angular coordinate since $g_{zz} \to 1$ as $\rho\to \infty$. Thus, our analysis on boost transformations at the asymptotic region still applies to this case as well.}
Similarly, although the extraordinary solutions belong to the tachyonlike string, some of them having $ |\mathfrak{p}|\geq 1/2\sqrt{3}$ are regular.
As explained above, there is no curvature singularity for $|\mathfrak{p}| \geq \sqrt{2}/{3}$ or $|\mathfrak{p}| = \sqrt{5/27}$.
Even if there appears a curvature singularity in the case of $1/(2\sqrt{3})\leq |\mathfrak{p}| < \sqrt{2}/{3}$ ($|\mathfrak{p}| \neq \sqrt{5/27}$), it is actually on the null hypersurface located at $\rho =K$ so that it does not affect physics outside.
These regular solutions might be useful in the future.
It would be of interest understanding how these regular geometries violating the gravitational dominant energy condition are actually formed through the collapse of matter.
Finally, it is interesting to see that tachyonlike string solutions might be possible since our metric ansatz assumes an infinitely extended or compactified cylindrical direction along $z$.
A formation of stationary solutions would be impossible if tachyonic matter collapse occurs with full spherical symmetry.

Finally, we point out that our analysis for stationary hypercylindrical solutions with momentum flow has been done only in five-dimensional spacetime.
However, we expect that the main results obtained in this paper still be valid qualitatively for such spacetime solutions in dimensions higher than five.



\begin{acknowledgments}

This work was supported by the Korea Research Foundation Grant
funded by the Korean Government(MOEHRD, Basic Research Promotion Fund), KRF-2008-314-C00063 and by the Topical Research Program of APCTP.
GK was also supported in part by Mid-career Researcher Program through NRF grant funded by the MEST (No. 2008-0061333) and by the Research Grant from SUN Micro Systems at KISTI.
HCK was also supported in part by the Korea Science and Engineering Foundation
(KOSEF) grant funded by the Korea government (MEST) (No.2010-0011308).
\end{acknowledgments}
 \vspace{1cm}



\end{document}